\documentstyle[prl,aps,preprint,amssymb,epsfig,floats]{revtex}
\tightenlines
\begin{document}

\title{Scaling of crossing probabilities for the $q$-state Potts model at
criticality}
\author{  O. A. Vasilyev}
\address{
 Landau Institute for Theoretical Physics, 142432 Chernogolovka, Russia
}

\maketitle

\begin{abstract}
 We present study of finite-size scaling and universality of crossing
probabilities for the $q$-state Potts model.  Crossing probabilities of the
Potts model are similar ones in percolation problem.  We numerically
investigated scaling of $\pi_{s}$ - the probability of a system to percolate
only in one direction for two-dimensional site percolation, the Ising model,
and the q-state Potts model for $q=3,4,5,6,8,10$.  We found the thermal
scaling index $y= \frac{1}{\nu}$ for $q<4$.  In contrast, $y \ne
\frac{1}{\nu}$ for $q=4$. 

\end{abstract}

\section{Introduction.}
In recent years the scaling and universality of the $q$-state Potts model
and percolation
are subject of intensive study~\cite{ZFA,LZ,BBFG,KL,dQ}.
There are two kinds of quantities of  interest
in  percolation problem: cluster distribution functions and 
crossing (spanning) probabilities.
With $n_{s}(p)$ - the number of clusters of size $s$ per lattice site, the
$m(p)= s_{max} n_{s_{max}}(p)=p ( 1- \sum \limits_{s \ne s_{max}} s
n_{s}(p) )$ is  probability of a lattice site to belong to "infinite"
cluster (the cluster of maximum size) and
 $\chi(p)= \sum \limits_{s \ne s_{max}} s^{2}
n_{s}(p)$ is the mean size of a finite cluster.
For an infinite lattice near the percolation point $p_{c}$ 
 $\chi(p) \sim
|p-p_{c}|^{-\gamma}$, $m\sim
(p-p_{c})^{\beta},\;\; p>p_{c}$~\cite{Stauffer}.
This quantities corresponds to magnetization $m(T)$ and magnetic
susceptibility $\chi(T)$
of the Potts model.

Critical properties of percolation may be found from the study of crossing
probabilities $\pi(p;L)$.
  Crossing probabilities were used by Reynolds, Stanley and
Klein for
renormalization group study of percolation~\cite{Rey1,Rey2}.
Universality of this crossing probabilities as 
functions of an aspect ratio $r$ in the critical point
was found in~\cite{Lang}.

The exact results for $\pi_{h}(r)$ and $\pi_{hv}(r)$  
in the percolation point was developed
by Cardy~\cite{Cardy} and Watts~\cite{Watts} for different $r$. 

Scaling properties of the function $\pi_{h}(p;L)$ - the probability,
 that a system
percolates in the horizontal direction for
two- and three-dimensional lattices was investigated by  Hu, Lin,
Chen~\cite{Hu1,Hu2,Hu3}.
They show, that $\pi_{h}(p)$   is  a universal function 
of scaling argument $x=(p-p_{c})L^{\frac{1}{\nu}}$, where $p_{c}$ - critical
point,
$L$ - lattice size, $\nu$ - correlation length exponent.

Universality 
of the crossing probability $\pi_{h}(\beta)$ for 
Ising model was investigated by Langlands et. all.~\cite{Langlands}.

In this paper we present  results of numerical investigations
of the crossing probability $\pi_{s}(\beta;L)$ (the probability of a system 
to percolate only in the one, horizontal either vertical, direction)
for the $q$-state Potts 
model.
We found,  that $\pi_{s}(\beta;L)$ is a universal function of 
the scaling variable $(\beta-\beta_{c})L^{y}$, where
for the Ising model and the $q=3$ Potts model the thermal scaling power 
$y=\frac{1}{\nu}$. Further, the limit of the crossing probability  at
the critical point $\pi_{s}(p_{c};L)$ is nonvanishing for the Ising model
and for the Potts model with $q=3,4$. In the case of  the Potts
model $q=5,6,8,10$
 percolation probabilities goes to zero, when the system size $L$
tends to infinity.

The crossing probability in the critical point behaves as 
$\pi_{s}(\beta_{c};L)\simeq A-a\log(L)$ for $L < \xi_{c}$, while
the  dependence is exponential $\pi_{s}(\beta_{c};L)\simeq \tilde A
\exp( -\tilde a L)$ for $L>\xi_{c}$, where $\xi_{c}$ is the
correlation
length in the critical point.

\section{Crossing probabilities.}

\subsection{Crossing probabilities for percolation.}

 Let us consider the site percolation on the square lattice.
Each site is occupied with probability $p$ and is empty with probability
$1-p$.
Let us  denote  by $\omega$ the sample, i.e.   the fixed distribution of
occupied
sites on the
lattice. The full set of samples $\Omega$ consists of $2^{L^{2}}$
configurations.

Let us introduce  indicator functions 
of crossing $I_{h}(\omega)$ in horizontal and $I_{v}(\omega)$ in vertical
directions~\cite{SK}.
 For example, if the cluster spans the sample  horizontally
 then $I_{h}(\omega)=1$ 
otherwise $I_{h}(\omega)=0$. 
Let us $S(\omega)$ - the number of occupied sites in the configuration
$\omega$.
 
The crossing probabilities for the site percolation on the square
lattice are introduced by 

\begin{equation}
\label{eqph}
\pi_{h}(p;L)=\sum \limits_{\omega \in \Omega}
P_{site}(\omega)I_{h}(\omega)=
\sum \limits_{\omega \in \Omega}
p^{S(\omega)}(1-p)^{L^2-S(\omega)}I_{h}(\omega)  
\end{equation}
\begin{equation}
\label{eqpv}
\pi_{v}(p;L)=\sum \limits_{\omega \in \Omega}
P_{site}(\omega)I_{v}(\omega)=
\sum \limits_{\omega \in \Omega}
p^{S(\omega)}(1-p)^{L^2-S(\omega)}I_{v}(\omega)  
\end{equation}

\begin{equation}
\label{eqphv}
\pi_{hv}(p;L)=\sum \limits_{\omega }
P_{site}(\omega)I_{h}(\omega)I_{v}(\omega)=
\sum \limits_{\omega \in \Omega}
p^{S(\omega)}(1-p)^{L^2-S(\omega)}I_{h}I_{v}(\omega)  
\end{equation}

\begin{equation}
\label{eqps}
\pi_{s}(p;L)=
\sum \limits_{\omega \in \Omega}
p^{S(\omega)}(1-p)^{L^2-S(\omega)} \left[
I_{h}(\omega)(1-I_{v}(\omega)) 
+(1-I_{h}(\omega)I_{v}(\omega) \right]  
\end{equation}

The function $\pi_{h}(p;L)$, defined by expression ~(\ref{eqph}),
is the probability, that the lattice with linear size $L$ percolates in the
horizontal
direction when the occupation probability is $p$. The function
$\pi_{v}(p;L)$
is  the probability, that the lattice percolates in the vertical
direction, $\pi_{hv}(p;L)$ is the probability, that the lattice 
percolates in both  directions,  
$\pi_{s}(p;L)$ is the probability, that the lattice 
percolates in the only direction (horizontal or vertical).
  It is clear, 
that on the square lattice $\pi_{h}(p;L)=\pi_{v}(p;L)$ and
 $\pi_{h}(p;L)=\frac{1}{2} \pi_{s}(p;L)+\pi_{hv}(p;L)$.
The example of functions $\pi_{s}(p;L)$ for different $L$ is shown on
the Fig~\ref{scfig}~a). 
 
\subsection{The $q$-state Potts model as a correlated site-bond
percolation.}

In the $q$-state Potts model the spin variable $\sigma$ takes values
from the set $\{1,2,\dots,q\}$. The probability $P(\omega)$ of  spin
configuration $\omega$ is defined by 
\begin{equation}
P_{Potts}(\omega)= \exp(-\beta
H(\omega))/Z(\beta),\;\;\; Z(\beta)=\sum \limits_{\omega}
P_{Potts}(\omega),\;\;
H= - J \sum \limits_{<i,j>}  \delta_{\sigma_{i},\sigma_{j}},
\label{gamp}
\end{equation}
where $H(\omega)$ - the Hamiltonian of the Potts model, $Z(\beta)$ is the
partition function, and $<i,j>$ is the sum over all neighbor sites.
We put through this paper the coupling 
constant $J=1$ and Boltzman factor $k_{B}=1$.

Fortuin and Kastelleyn propose  mapping of $q$-state Potts model
onto
the site-bond correlated percolation~\cite{FK}. It was used by Swendsen and
Wang
for their Monte-Carlo cluster algorithm~\cite{SW}. The single-cluster
algorithm was proposed by Wolff~\cite{Wolff}. Thermodynamical quantities
can be expressed  in terms of clusters in correlated site-bond 
percolation~\cite{Coniglio,Onorio}.
  
Lets the bond between neighbor sites with  same values of spin variable
$\sigma_{i}=\sigma_{j}$ is closed $b_{i,j}=1$ with probability
$r=1-\exp(-\beta)$ and opened $b_{i,j}=0$  with probability $1-r$, and the
bond between neighbor sites with different values of spin variables
is always opened.

Define the chosen configuration of closed and opened bonds by $\upsilon$.
Then joint distribution of the spin configuration $\omega$ and the bond
configuration $\upsilon$ is (see\cite{ES,Onorio}) 

\begin{equation}
\label{defp}
P_{Potts}(\omega,\upsilon)= Z^{-1}(\beta) \prod \limits_{<i,j>}
\left[  (1-r)
\delta_{b_{i,j},0}+ r \delta_{\sigma_{i},\sigma_{j}}
\delta_{b_{i,j},1}\right],\;\;\; Z(\beta)=\sum \limits_{\omega}
\sum \limits_{\upsilon } P(\omega,\upsilon)
\end{equation}

where $P_{Potts}(\omega)=\sum \limits_{\upsilon} P_{Potts}(\omega,\upsilon)$
is the probability of spin configuration $\omega$ 
 and $Z(\beta)$ is partition function for $q$-state Potts model.
To each spin configuration $\omega$ in accordance with~(\ref{defp}) 
corresponds a bond configuration $\upsilon$, when bonds between 
 neighbor sites with different values of spin variables are always opened,
and the bonds between sites with equal values of spin variables
are closed with probability $r=1-\exp(-\beta)$ and opened with probability 
$1-r$. The subset of sites, connected by closed bonds, called 
"physical" cluster. If we  assign to spin variables in  each "physical"
cluster the random value from the set $\{1,2,\dots,q\}$,  we obtain the   
new spin configuration. This procedure describes the Swendsen-Wang cluster
algorithm, used in this paper. 

\subsection{Crossing probabilities for the $q$-state Potts model.}

\begin{figure}  
\vbox{
\setlength{\unitlength}{0.240900pt}
\ifx\plotpoint\undefined\newsavebox{\plotpoint}\fi
\begin{picture}(1,1)(0,0)
\end{picture}
\hspace{-5mm}\mbox{
\mbox{
\epsfxsize=75mm 
\epsfysize=75mm
\epsfbox{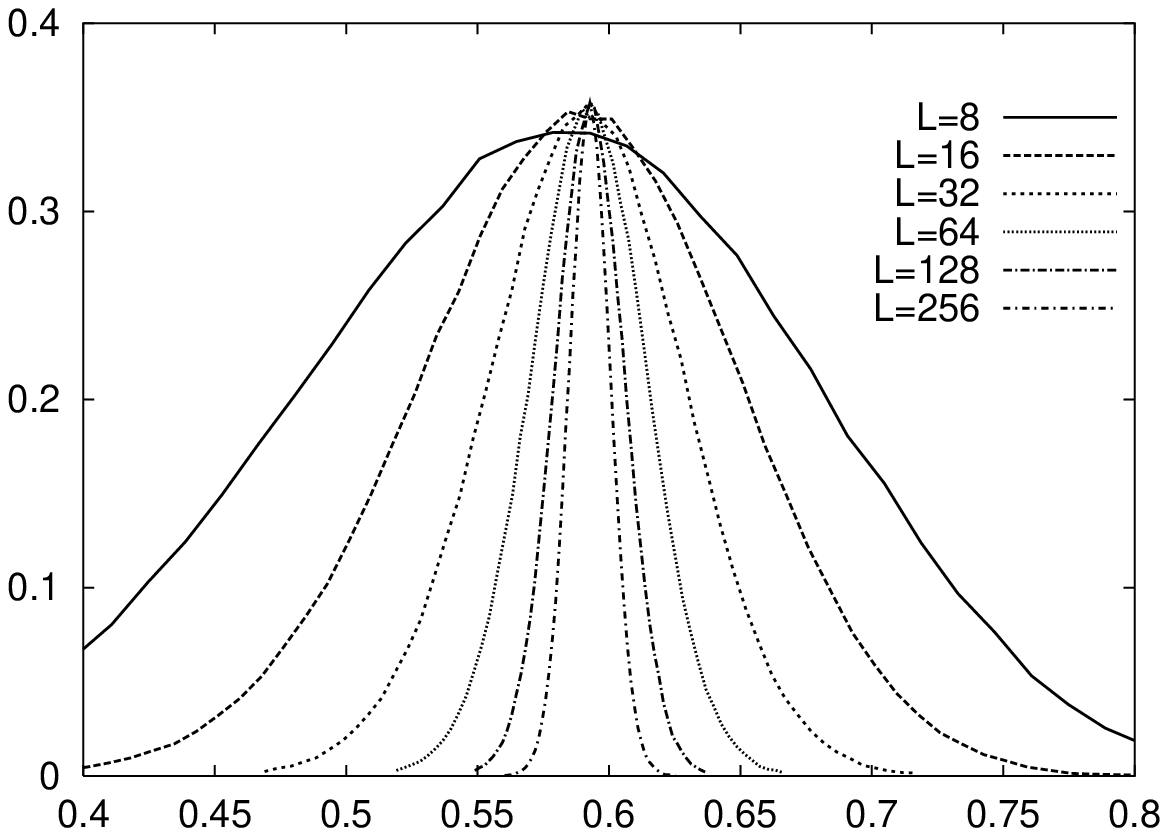}
}
\mbox{
\epsfxsize=75mm
\epsfysize=75mm
\epsfbox{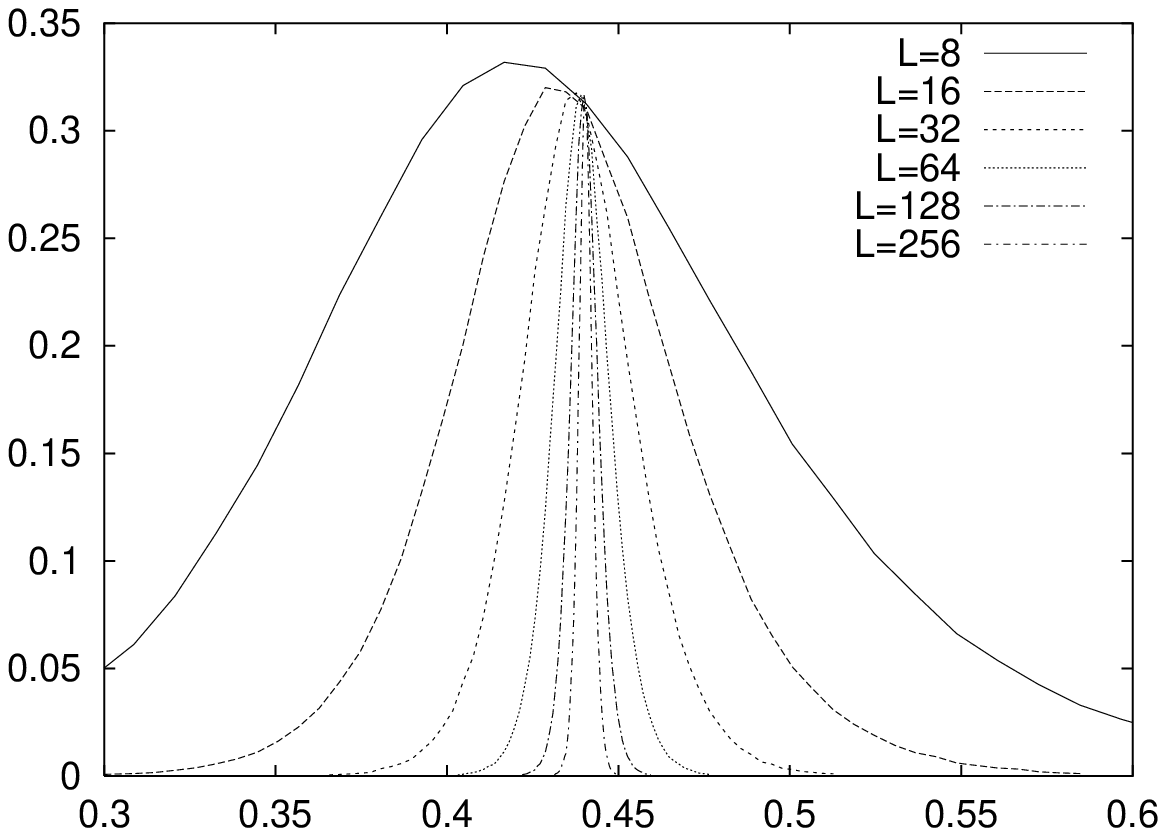}
}}%
\begin{picture}(0,0)
\put(-1900,495){\small  $ \pi_{s}(p;L)$}
\put(-1370,-20){\large $p$}
\put(-960,450){\small $\pi_{s}(\beta;L)$}
\put(-440,-20){\large $\beta$}
\put(-1850,800){\large\bf a)}
\put(-930,800){\large\bf b)}
\end{picture}

\hspace{-5mm}\mbox{
\mbox{
\epsfxsize=75mm 
\epsfysize=75mm
\epsfbox{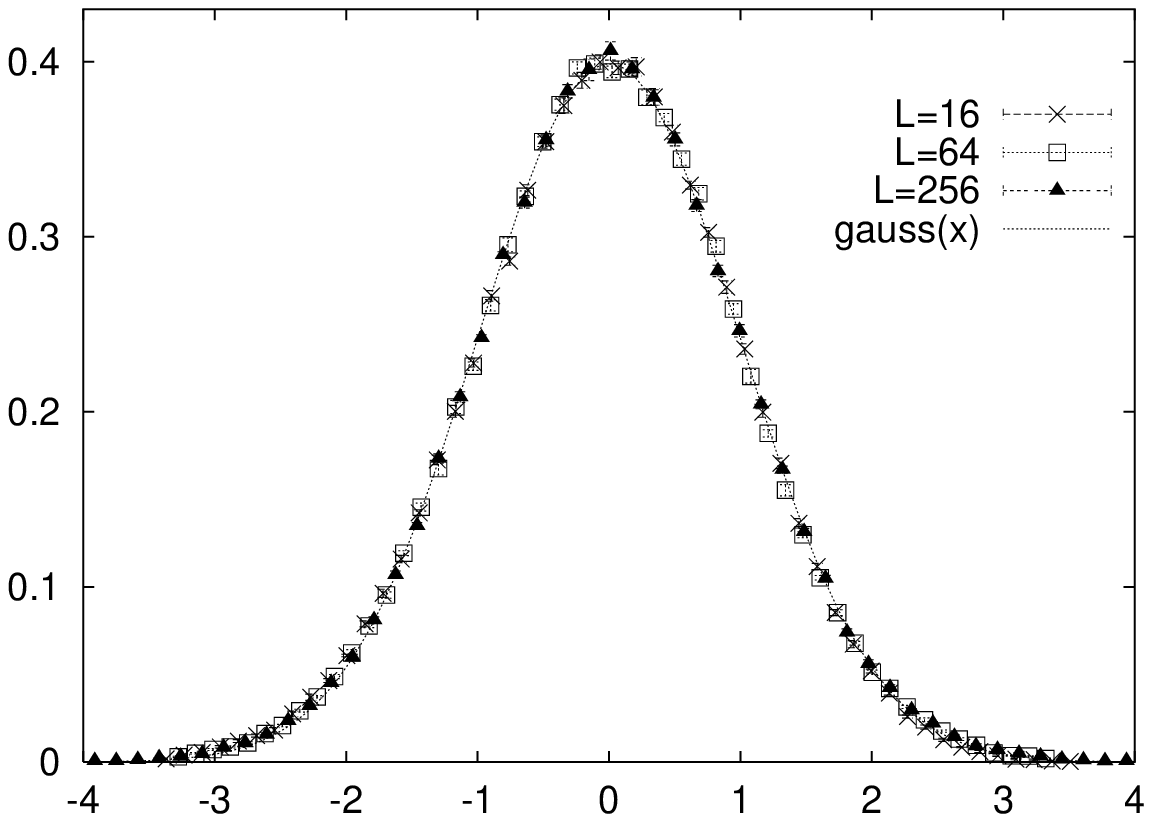}
}
\mbox{
\epsfxsize=75mm
\epsfysize=75mm
\epsfbox{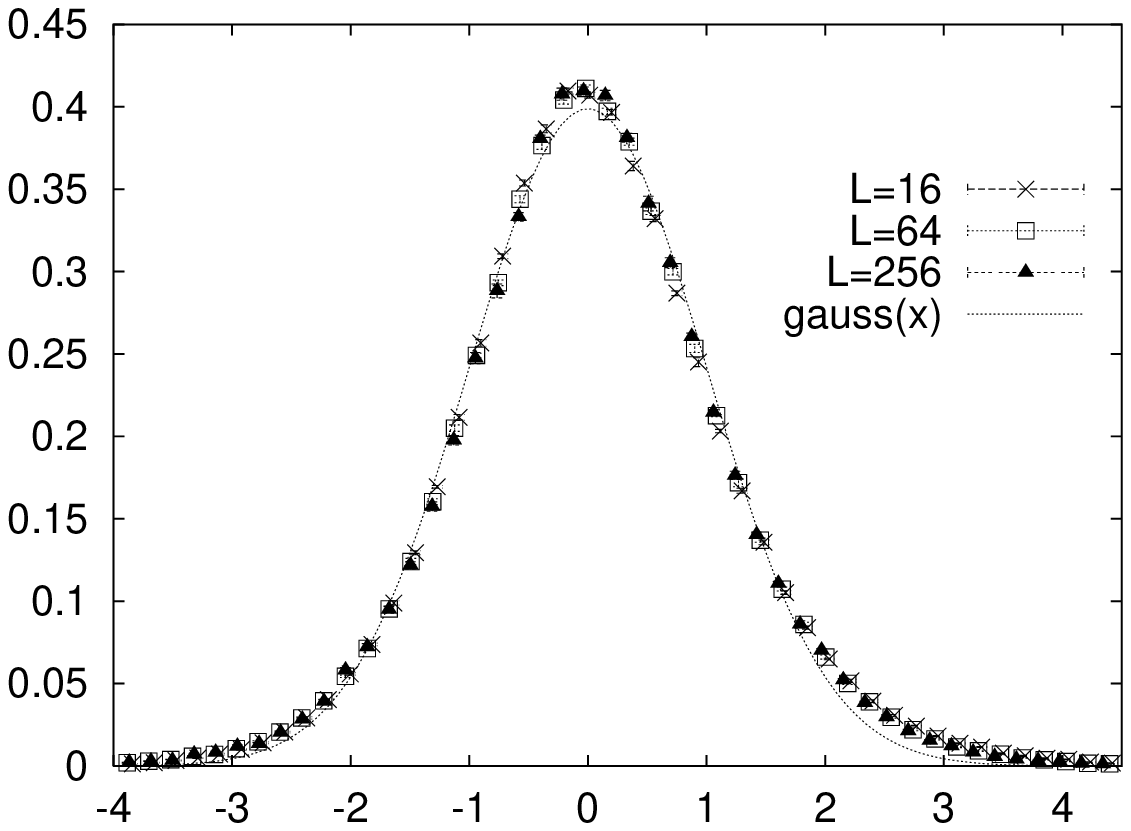}
}}%
\begin{picture}(0,0)
\put(-1900,460){\small  $ \tilde\pi_{s}(x)$}
\put(-1365,-20){\large $x$}
\put(-940,450){\small $\tilde \pi_{s}(x)$}
\put(-440,-20){\large $x$}
\put(-1850,800){\large\bf c)}
\put(-930,800){\large\bf d)}
\end{picture}
\caption{Crossing probabilities for percolation and Ising model.}
  $\pi_{s}(p;L)$   for percolation (a) and  for the Ising model (b)\\   
$\tilde \pi_{s}(x)$ - normalized crossing probabilities  for percolation (c)
and   for the Ising model (d)
\label{scfig}
}
\end{figure}
For  the percolation problem the crossing probabilities 
$\pi_{h}(p;L)$, $\pi_{v}(p;L)$, $\pi_{hv}(p;L)$, $\pi_{s}(p;L)$
are defined 
 as indicator functions $I_{h}(\omega) $, $ I_{v}(\omega)$
 averaged over all site (or bond) 
configurations in accordance with
~(\ref{eqph}),~(\ref{eqpv}),~(\ref{eqphv}),~(\ref{eqps})
Crossing probabilities are probabilities  for a system with linear size  $L$
to percolate 
according chosen rules at probability $p$ of a site (or bond) to be
occupied.

For a $q$-state Potts model  statistical weights of site-bond configuration
(in terms of Fortuin-Kasteleyn mapping)
are defined by expressions ~(\ref{defp}),
therefore it is  natural to define the crossing probabilities
through the indicator functions $I_{h}(\omega,\upsilon)$,
$I_{v}(\omega,\upsilon)$ 
averaged over all site and
bond configurations.

\begin{equation}
\label{defpi}
\pi_{h}(\beta;L)=\sum \limits_{\omega }
\sum \limits_{\upsilon }
P_{Potts}(\omega,\upsilon)I_{h}(\omega,\upsilon) 
\end{equation}
where $\pi_{h}(\beta;L)$  is a probability, that at least one percolating in
the horizontal direction  
"physical" cluster exists, $\beta=\frac{1}{T}$ is the inverse temperature.
 By the same way others crossing probabilities
are defined.

\section{Scaling of crossing probability function $\pi_{s}(\beta;L)$
for the two-dimensional $q$-state Potts model.} 
\label{secscal}

\subsection{Features of approximation.}
\label{secdet}

We use the Swendsen-Wang algorithm to generate the different
spin configurations. For each spin configuration
we generate the bond configuration. Then we decompose the lattice into 
independent clusters of connected sites using Hoshen-Kopelman~\cite{HK}
algorithm. After that we analyze crossing  properties
$I_{h}(\omega,\upsilon)$,
$I_{v}(\omega,\upsilon)$ of this configuration.  
We average indicator functions $I_{h}(\omega,\upsilon)$,
$I_{v}(\omega,\upsilon)$ over
$N$ configurations.
\begin{equation}
\label{avrpi}
\pi_{h}(\beta)=\frac{1}{N} \sum \limits_{l=1}^{N} I_{h}(\omega_{l}),\;\; 
\pi_{s}(\beta)=\frac{1}{N} \sum \limits_{l=1}^{N}
\left[I_{h}(\omega_{l})\left(1-I_{v}(\omega_{l}) \right) 
+I_{v}(\omega_{l})\left( 1- I_{h}(\omega_{l})\right)
\right]
\end{equation}

We compute crossing probabilities for $40-50$
values of $\beta$ in the interval $[\beta_{c}- d \beta(L),\beta_{c}+ d
\beta(L)]$, where the width of the interval was proportional 
$d \beta \sim L^{-y}$ - see~Fig.~\ref{scfig}~b). The total number of 
configurations for every value of $\beta$ are  $N=2\times 10^{4}-15\times
10^{4}$. It should be noted, that $\beta_{c}(q=2)=
\log(\sqrt{2}+1)=2 \beta_{c}(Ising)=0.8813735870...$.

 For percolation 
we generate site configurations by "grand canonical" method~\cite{GL} - 
every  site is occupied with probability $p$ and is empty with probability
$1-p$.
 For each concentration $p$ we
generate
$N=5\times 10^5$ configurations. 
The number of analyzed configurations is the same 
for small and large lattice sizes.
  We use four-point shift-register random number generator
with maximum length 9689~\cite{Shchur}.
 The  crossing probabilities are not self-averaging by
definition~(\ref{avrpi}),
and the variance of $\pi_{h}$ and $\pi_{s}$ not depend upon lattice size.

Then we approximate crossing probabilities $\pi_{s}(\beta;L)$ for different
$q$ by the Gauss function 
\begin{equation}
\label{gauss}
\pi_{s}(L,\beta) \simeq A(L) \exp \left( - \frac{1}{2}(\beta -
\beta_{c}(L))^{2} B^{2}(L) \right)
\end{equation}
  
Normalized crossing probabilities $\tilde \pi_{s}(x)$ looks
like Gaussian
 - see.~Fig.~\ref{scfig}~c)-d). On this figures we placed normalized 
crossing probabilities $\tilde \pi_{s}(x) =\frac{1}{\sqrt{2 \pi} A(L)}
\pi_{s}( (\beta-\beta_{c}(L))B(L);L)$ 
where $x=(\beta-\beta_{c}(L))B(L)$.
For normalization we use parameters $A(L)$, $B(L)$, $\beta_{c}(L)$,
which we obtained as a result of approximation of the numerical data by
the expression~(\ref{gauss}).

We dont approve, that the crossing probability is Gauss function,
although they look similar~Fig.~\ref{scfig}~c),~Fig.~\ref{scfig}~d).
We use this approximation like convenient method to compute the amplitude,
the inverse variance and the maximum location
of this function.

As a result of approximations, for every investigated value of $q$ and $L$ 
we get  the set of values $A(L)$, $B(L)$ É $\beta_{c}(L)$. 

\subsection{Scaling of crossing probabilities.}

\label{secapp}

\begin{figure}[h]
\epsfxsize=100mm
\epsfysize=80mm  
\centerline{\epsfbox{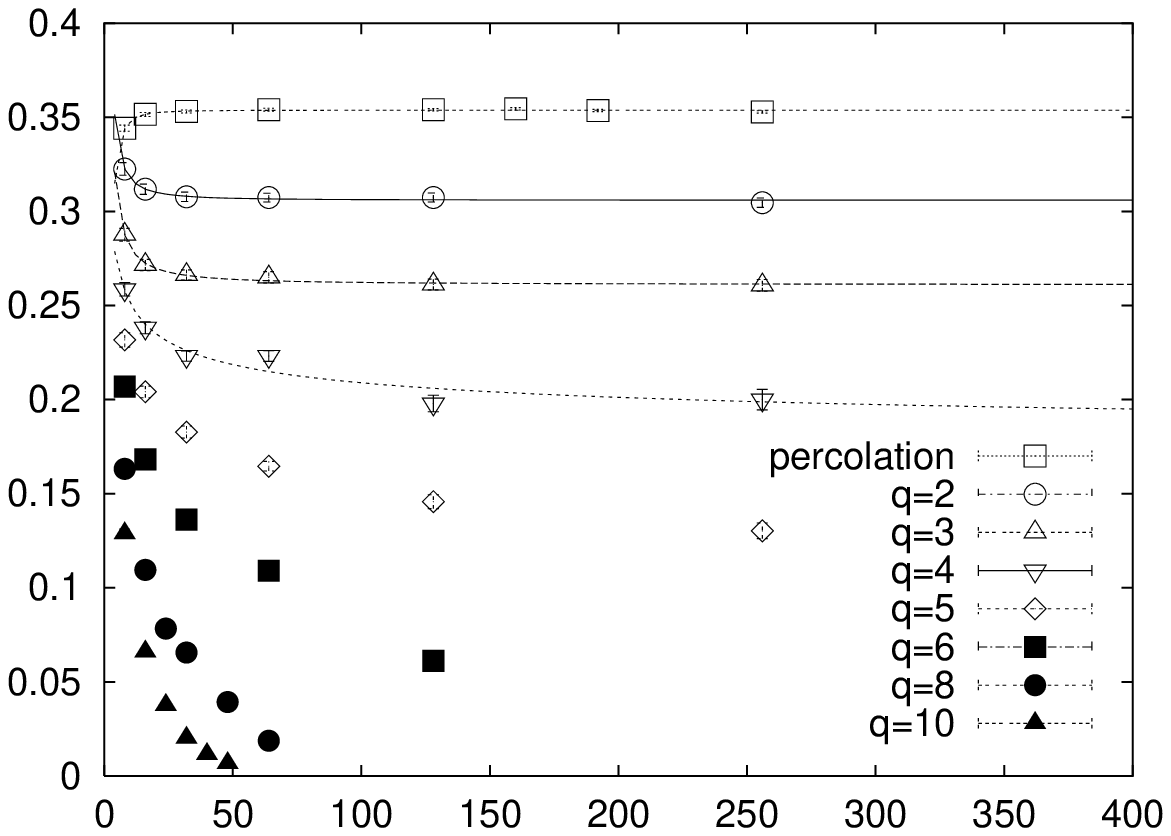}}
\begin{picture}(0,0)
\setlength{\unitlength}{1.000pt}
\put(55,+140){ $A(L)$}
\put(+210,20){ $L$}
\end{picture}
\caption{ Amplitudes $A(L;q)$ of  crossing probabilities
$\pi_{s}(\beta;L)$.} 
\label{qa1fig}
\end{figure}

We plot $A$ as a function of $L$ on Fig.~\ref{qa1fig}.
For $q\le 4$  we expect finite size corrections to 
limit values in the form $a L^{-x}$, as it was shown for
percolation~\cite{Ziff}. 
We found, that the functions $A(L)$ for $q=2,3,4$ could be well approximated 
by
\begin{equation}
\label{fora}
A(L)\simeq 
A_{0} +a L^{-x}, \;\; q\le 4
\end{equation}
Results of approximation
plotted on Fig.~\ref{qa1fig} by lines.
 Second, third and fourth rows of the
Table~\ref{appr} represents result for amplitude $A_{0}$, and scaling terms
$a$ and $x$ for $q\le 4$. In the Table~\ref{appr} below a row
with values of $q$ placed results of approximation for this $q$.
The case $q=4$ is intermediate of models with phase transitions of the 
first order
$q>4$ and the second order $q\le 4$~\cite{Baxter}.
 
\begin{figure}[h]
\epsfxsize=100mm
\epsfysize=80mm  
\centerline{\epsfbox{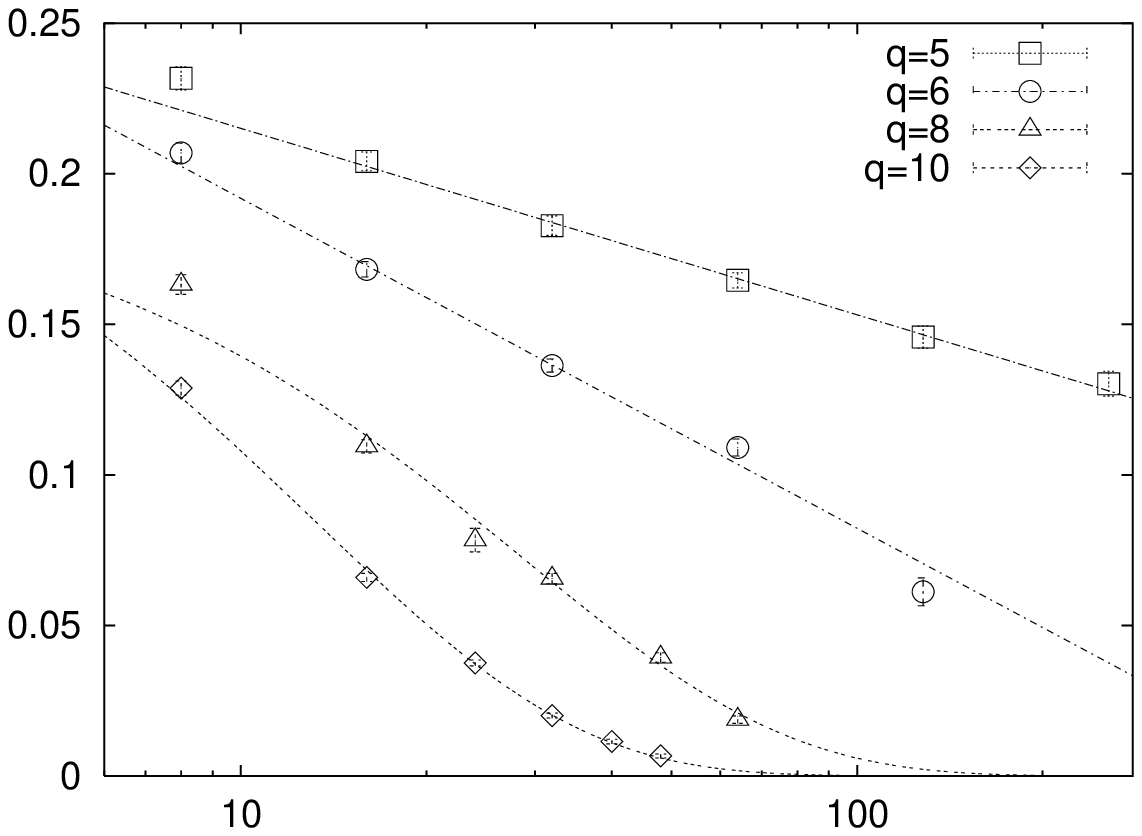}}
\begin{picture}(0,0)
\setlength{\unitlength}{1.000pt}
\put(55,+140){ $A(L)$}
\put(+210,20){ $L$}
\end{picture}
\caption{ Amplitudes $A(L;q)$ of  crossing probabilities
 $\pi_{s}(\beta;L)$ in $\log(L)$ scale.}
\label{qafig}
\end{figure}

\begin{figure}[h]
\epsfxsize=100mm
\epsfysize=80mm  
\centerline{\epsfbox{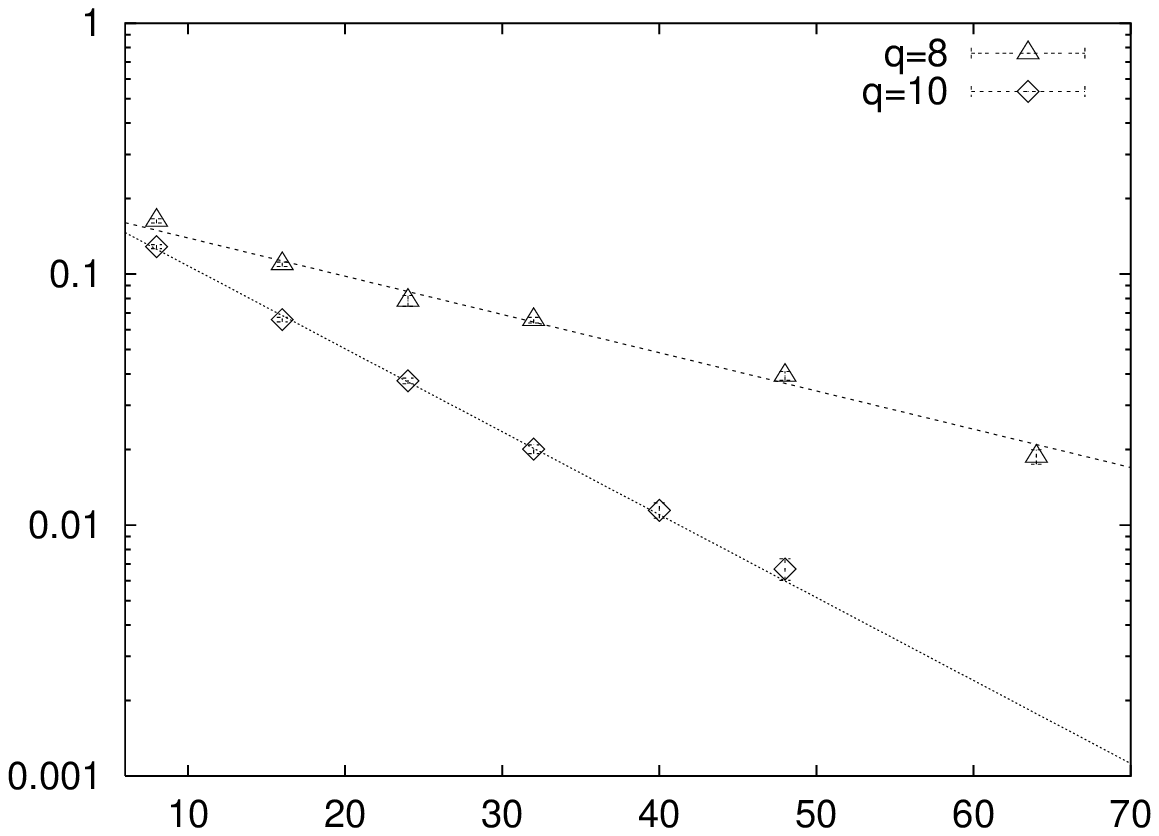}}
\begin{picture}(0,0)
\setlength{\unitlength}{1.000pt}
\put(55,+150){ $A(L)$}
\put(+210,20){ $L$}
\end{picture}
\caption{Amplitudes $A(L;q)$ of  crossing probabilities
 $\pi_{s}(\beta;L)$
  in $\log(A)$ scale.}
\label{qaefig}
\end{figure}
We can expect, that for $q>4$ the crossing probability tends to zero with
$L$ increasing, because the correlation length in the critical 
point $\xi_{c}$ is finite~\cite{Arisue}.
 Values of correlation length   $\xi_{c}$ for
$q>4$ in sixth row of Table~\ref{appr} are  taken from~\cite{Arisue}, 
 We can also expect the different
behavior of
$A(L)$ 
 for cases $L< \xi_{c}$ and $L>\xi_{c}$
For $q=5,6,8,10$  behavior
of $A(L)$ have another form - see~Fig.~\ref{qafig}.

\begin{equation}
\label{foral}
A(L)\simeq 
A_{0}-a \log(L),\;\; q >4,\;\; L <\xi_{c} 
\end{equation}

\begin{equation}
\label{forae}
A(L)\simeq 
A_{0} \exp(- a L),\;\; q >4,\;\; L >\xi_{c} 
\end{equation}

On this figure we use the logarithmic scale for $L$ axis, and we see, that
points for $q=5,6$ lay on  straight lines. So we approximate data for 
$q=5,6$ by the logarithmic low~(\ref{foral}). We put results of
approximation
into seventh and eighth rows of the Table~\ref{appr}. Also we plote 
results of approximation on
the Fig.~\ref{qafig} by lines.

\begin{figure}[h]
\epsfxsize=100mm
\epsfysize=80mm  
\centerline{\epsfbox{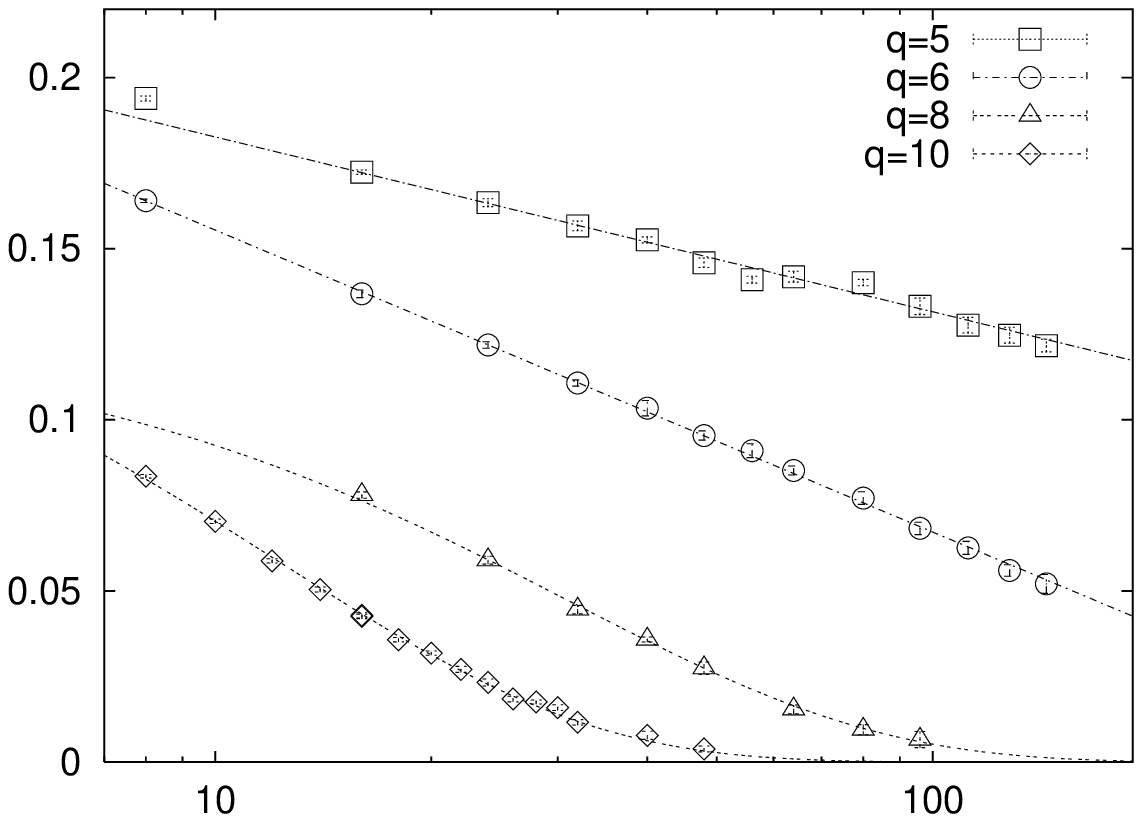}}
\begin{picture}(0,0)
\setlength{\unitlength}{1.000pt}
\put(60,+147){ $\pi_{s}(\beta_{c};L)$}
\put(+210,20){ $L$}
\end{picture}
\caption{ Crossing probability at  critical points 
 $\pi_{s}(\beta_{c};L)$  in $\log(L)$ scale.}
\label{dqafig}
\end{figure}
\begin{figure}[h]
\epsfxsize=100mm
\epsfysize=80mm  
\centerline{\epsfbox{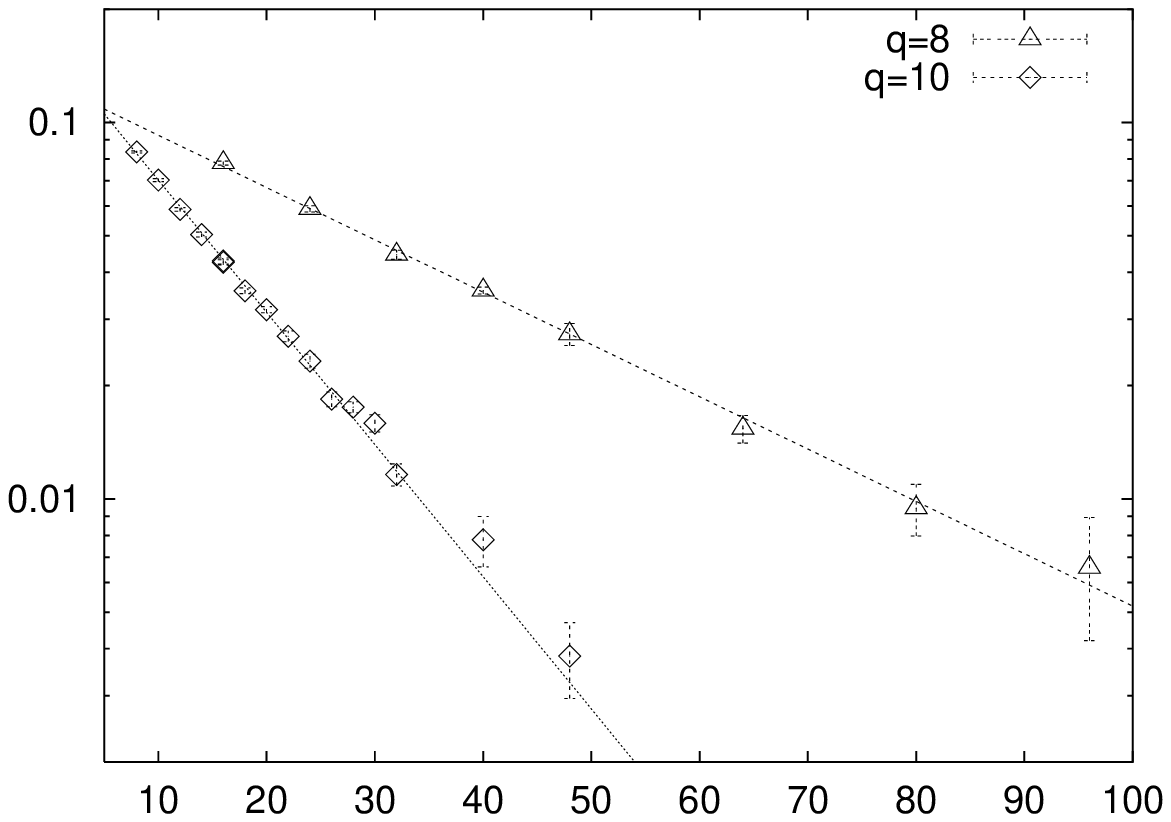}}
\begin{picture}(0,0)
\setlength{\unitlength}{1.000pt}
\put(60,+140){ $\pi_{s}(\beta_{c};L)$}
\put(+210,20){ $L$}
\end{picture}
\caption{ Crossing probability at  critical points
 $\pi_{s}(\beta_{c};L)$  in $\log(\pi_{s})$ scale.}
\label{dqaefig}
\end{figure}

But the data for $q=8,10$ looks like  straight lines, when we use 
the logarithmic scale for $A(L)$ axis, while the $L$ axis
use the normal scale - see~Fig.~\ref{qaefig}. Therefore we approximate
$A(L)$
for $q=8,10$ by the exponential  low~(\ref{forae}).  
Results of approximation for $q=8,10$ by
the exponential formula~(\ref{forae}) are placed in seventh and eights
rows of Table~\ref{appr} and are marked by stars $*$, 
and plotted on Fig.~\ref{qafig},~Fig.~\ref{qaefig} by lines. 

We see, that for $q=5,6$  the correlation length is greater, 
then the linear size of our lattices $L=8,\dots,256$.
But for the case $q=10$ we examine region $L>\xi_{c}$.
For $q=8$ the value $\xi_{c}=23.8$ and only two sizes $L=8,16$
are smaller, than $\xi_{c}$. 

We make additional calculations to check this approximation
formulas~(\ref{foral})~(\ref{forae}).
We compute  the crossing probability $\pi_{s}(\beta;L)$ for $q=5,6,8,10$
 direct in  critical points $\beta_{c}(q)=\log(\sqrt(q)+1)$~\cite{Baxter}.
Results of this calculation presented on Fig.~\ref{dqafig} and
Fig.~\ref{dqaefig}. This figures corresponded to figures Fig.~\ref{qafig}, 
Fig.~\ref{qaefig} respectively.
 
  The picture, which we can see in this
figures, allowed  us to make the
following conclusion: for the case $q>4$ the amplitude $A(L)$ of crossing
probability goes to
zero by the logarithmic low~(\ref{foral}) - see Fig.~\ref{qafig},
Fig.~\ref{dqafig}, when $L<\xi_{c}$ 
goes  by the exponential low~(\ref{forae}) - see Fig.~\ref{qaefig},
Fig.~\ref{dqaefig}, when $L>\xi_{c}$.   
For the case $q\le 4$ the amplitude $A(L)$ goes to finite value
by the power low~(\ref{fora}).

\begin{figure}[h]
\epsfxsize=100mm
\epsfysize=80mm  
\centerline{\epsfbox{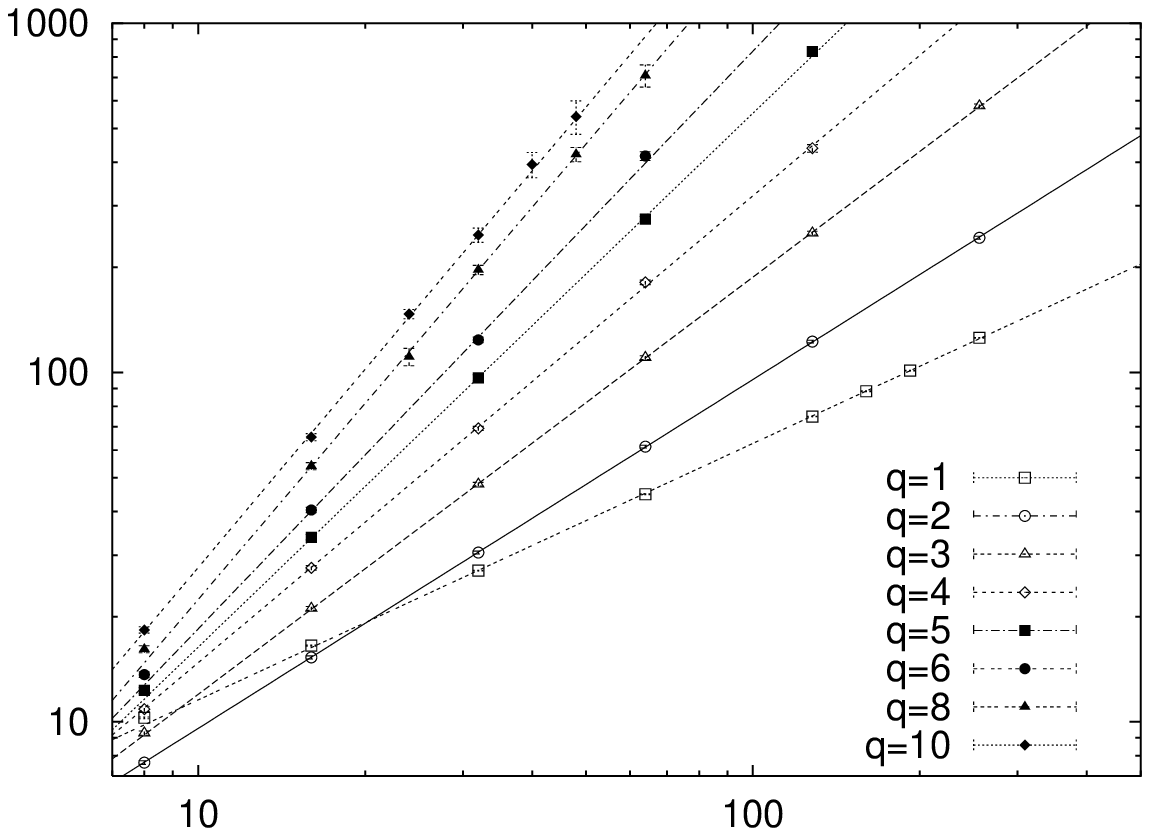}}
\begin{picture}(0,0)
\setlength{\unitlength}{1.000pt}
\put(55,+165){ $B(L;q)$}
\put(+210,20){ $L$}
\end{picture}
\caption{The inverse width $B(L)=\frac{1}{W} (L)$ of crossing
probabilities.} 
\label{qwfig}
\end{figure}

The expression for the inverse dispersion $B$ in~(\ref{gauss}) follows from
the scaling
relation (for the case of the Gauss function~(\ref{gauss}), parameter $B$
is the inverse dispersion $B=\frac{1}{W}$). 
The universality of the function $\pi_{s}(\beta;L)$ implies, that
the inverse dispersion is proportional the lattice size $L$ in the power
of $y=\frac{1}{\nu}$ at least for  cases of the second order phase
transitions,
when the critical index $\nu$ of correlation length is defined - see 
formula~(\ref{forb}).
Behavior of parameter $B(L)$ as a function of linear size $L$ is shown on
the Fig.~\ref{qwfig} in  log-log scales. The points for different
$q$ lay on  straight lines in accordance with~(\ref{forb}) 

\begin{equation}
\label{forb}
\frac{1}{W(L)} = B(L) \simeq b L^{y}
\end{equation}

In eleventh, twelfth, fourteenth, fifteenth rows Table~\ref{appr} placed
results for $b$ and $y$ of approximation of inverse 
width $B(L)$~(\ref{forb}). In the tenth row we put analytical
values of $\frac{1}{\nu}$ for $q\le 4$. 
 We see, that for $q<4$ the scaling parameter
$y$ is equal the inverse critical index $\nu$ in agreement with
scaling relations $\pi_{s}(\beta;L)=
f(L^{\frac{1}{\nu}}(\beta-\beta_{c}(L)))$.
But for $q=4$ we see, that $y \ne \frac{1}{\nu}$. 
For $q>4$ we obtain values of thermal scaling exponents $y$,
but it is unclear, how to relate it to another critical
exponents.

For the locations $\beta_{c}(L)$, maximum of the function
$\pi_{s}(\beta;L)$,
we expect
deviations from  limit values in the power form $L^{-z}$. For the
percolation,
the power $z$ seems to be equal $z=1+\frac{1}{\nu}=1.75$~\cite{Ziff}.
But the same deviation of maxima for magnetic susceptibility and specific
heat of the
Ising model is proportional to $L^{\frac{1}{\nu}}$~\cite{FF,Landau}
So we considered $z$ as a free parameter and use for approximation
formula~(\ref{forc}). 

\begin{equation}
\label{forc}
\beta_{c}(L) \simeq \beta_{c}(\infty) +c L^{-z}
\end{equation}

In seventeenth and twenty second rows of the Table~\ref{appr} we put the
precise values of 
critical points. For $q=2,3,4,5,6,8,10$ we use expression
$\beta_{c}(q)=\log(\sqrt{q}+1)$.
We use data from~\cite{Ziff}  for the critical point for
percolation. 
 
The position of the maximum of the crossing probability
$\pi_{Ps}(\beta;L)$ on the lattice with linear size $L$ goes 
to the critical point by power low~(\ref{forc}). We placed
parameters of approximation $\beta_{c}(\infty)$, $c$, $z$ in eighteenth,
nineteenth, twentieth, twenty-third, twenty-fourth and twenty-fifth rows
in the Table~\ref{appr}. 

There in good agreement between results of approximation and 
analytical values of critical temperature. This values deviates
only in fifth and fourth digits  after the decimal point. 

\section{Summary.}

\begin{itemize}

\item crossing probabilities for $q$-state Potts model are
universal functions of scaling variable
$(\beta-\beta_{c})L^{\frac{1}{\nu}}$,
looks like  the Gauss function $A \exp(-  (\beta-\beta_{c}(L))^{2}B^{2}
L^{\frac{2}{\nu}})$.

\item  locations of  maxima of crossing probabilities
$\beta{c}(L;q)$ goes to phase transition points
$\beta(\infty;q)=\log(\sqrt{q}+1)$ when $L$ goes to $\infty$.

\item for the case $q<4$ the thermal scaling index $y=\frac{1}{\nu}$,
where $\nu$ is the critical index of correlation length.  For
$q=4$ the thermal index $y\ne \frac{1}{\nu}$. For $q \le 4$ the limit of the
crossing
probability in the critical point is nonzero. 

\item for $q>4$ crossing probabilities in  critical points
go to zero by the logarithmic low, when $L<\xi_{c}$ and by the exponential
low, when $L>\xi_{c}$, where $\xi_{c}$ is the correlation length in the critical
point.

\end{itemize}

\section{Acknowledgments}
Author is thankful to L. N. Shchur for a kind discussion and help. 
 Author is grateful to Landau stipendium committee
(Forshungzentrum/KFA, J\"ulich) for support. 

\begin{table}
\caption{Results of approximation by (\ref{fora})-(\ref{forc}).}
\vbox{
\begin{tabular}{|c|c|l|l|l|l|}
1& $q$&  $q=1 $ & $q=2 $ &  $q=3$ & $q=4 $\\
\hline
2& $A_{0}$   & 0.3538(3)& 0.3059(8) & 0.2610(12)&
0.17(5)\\
\hline
3& $a$   & -0.73(60)& 0.36(23) & 0.32(13)&
0.17(3)\\
\hline
4& $x$  & -2.1(4) & 1.48(33) & 1.20(20)&
0.3220(29)\\
\hline
\hline
5& q&  $q=5 $ & $q=6 $ &  $q=8$ & $q=10 $\\
\hline
6& $\xi(\beta_{c})$ - see. \cite{Arisue} & 2512.1 & 158.9& 23.9 & 10.6\\
\hline
7& $A_{0}$   & 0.277(3)& 0.301(15) & $0.198(27)^{*}$& $0.231(7)^{*}$\\
\hline
8& $a$   & 0.0269(8) & 0.048(4) & $0.035(3)^{*}$&
$0.076(1)^{*}$\\
\hline
\hline
9& $q$ &  $q=1$ & $q=2$ & $q=3$ & $q=4$\\
\hline
10& $\frac{1}{\nu}$  &  0.75  &  1.0  &  1.2  &  1.5\\
\hline
11& $y$ & 0.7343(28) & 0.9988(19) & 1.1945(26) &
1.337(11)\\
\hline
12& $b$   & 2.131(29) & 0.961(8) & 0.766(8) &
0.68(3)\\
\hline 
\hline
13& q&  $q=5 $ & $q=6 $ &  $q=8$ & $q=10 $\\
\hline
14& $y$   & 1.5284(85) & 1.654(33) & 1.865(13)&1.886(19)\\
\hline
15& $b$   & 0.485(16)& 0.450(45) & 0.306(13)&0.360(18)\\
\hline
\hline
16& q &  $q=1$ & $q=2$ & $q=3$ & $q=4$\\
\hline
17& $\beta_{c}$ exact  & $p_{c}=0.592746$  &  $0.881374^{**}$  &  1.005053
 & 1.098612\\
\hline
18& $\beta_{c}(\infty)$   & $ p_{c}(\infty)=0.592731(28)$ & 0.881267(28) &
1.004990(19)
 & 1.098600(22)\\
\hline
19& $c$ & -0.333(12) & -0.309(17) & -0.527(24) & 
-0.74(4)\\
\hline
20& $z$ & 1.57(12) & 1.148(17) & 1.277(14) &
1.391(18)\\
\hline
\hline
21& q&  $q=5 $ & $q=6 $ &  $q=8$ & $q=10 $\\
\hline
22& $\beta_{c}$ exact &  1.174359& 1.238226 &  1.342454 & 1.426062\\
\hline
23& $\beta_{c}(\infty)$   & 1.17440(6)& 1.23828(5)&
1.3427(5)&
 1.42648(18)\\
\hline
24& $c$  & -1.114(24)& -1.30(9) & -1.5(8)&
-1.89(15)\\
\hline
25& $z$  & 1.542(7) & 1.608(25) & 1.68(19)&
1.78(3)\\
\end{tabular}

* - approximation by exponent - formula~(\ref{forae})\\
** - $\beta_{c}(q=2)=\log(\sqrt{2}+1)=2\beta(Ising)=0.881373587...$
}
\label{appr}
\end{table}

\end{document}